\documentclass{aa}
\usepackage{epsfig}

\begin{document}

\thesaurus{10(07.03.2 46P/Wirtanen)}  
\title {Oblate spheroid model of nucleus of Comet 46P/Wirtanen}
\author{Ma{\l}gorzata Kr{\'o}likowska \and S{\l}awomira Szutowicz}
%
\institute{Space Research Centre of the Polish Academy of Sciences,
           Bartycka 18A, 00-716 Warsaw, Poland}

\offprints{M. Kr{\'o}likowska, e-mail: mkr@cbk.waw.pl}
%
\date{Received  / Accepted 16 December 1998 }
   \maketitle
   \markboth{M.\,Kr{\'o}likowska \& S.\,Szutowicz: Model of 46P/Wirtanen}
{M.\,Kr{\'o}likowska \& S.\,Szutowicz: Model of 46P/Wirtanen.}

   \begin{abstract}
An improved forced-precession model of Comet 46P/Wirtanen is presented. The
nongravitational motion of the comet has been investigated based on the
forced precession model of the rotating cometary nucleus. The least squares method
applied to observational equations allows us to determine six basic
nongravitational parameters: $A$, $\eta$, $I$, $\phi$, $f_p$, and $s$ together
with six orbital elements. The solutions were obtained with additional
assumptions: 1. The cometary activity with respect to the perihelion is asymmetric and,
2a. The time shift parameter which describes the displacement of maximum activity
with respect to the perihelion could change its value during the investigated
apparitions or 2b. The real activity of the comet changed during the time
interval considered . All eight recorded apparitions of the comet were linked using all
astrometric observations covering the period 1948--1997 with a mean RMS
residual of 1\farcs 6. According to the best solution, the nucleus of Comet
Wirtanen is oblate along the spin-axis with a ratio of equatorial to polar radius of
about $R_a/R_b=1.1$. This precession model yields a value of 4.9~hrs/km for
the ratio of $P_{\mathrm{rot}}/R_a$. Assuming that the nucleus radius of 46P/Wirtanen most
likely is in the range of 0.5--2.0~km we obtain a range 2.5 -- 10.0 hours for
the rotational period.
\end{abstract}

\section{Introduction}

\noindent The short-period comet 46P/Wirtanen will be investigated for two years
from 2011 during the ROSETTA cometary rendez-vous mission. Therefore, precise
predictions of the future perihelion passages and space positions of the comet
are crucial for successful planning of the mission.

Comet 46P/Wirtanen has been observed at eight returns during the almost fifty-year
interval since its discovery in 1948 by C.A.~Wirtanen. Two close approaches of
the comet to Jupiter in 1972 and 1984 are responsible for significant changes
of orbital elements. Apart from the planetary perturbations the comet is
subject to nongravitational forces (\cite{Mar2} 1997). The method of determining
the nongravitational effects in the orbital motion of the comet was proposed by
\cite{Mar1} (1973). In Marsden's formalism the three orbital components of
a nongravitational force acting on the comet have the form:
\begin{equation}
          F_i = A_ig(r), ~~~~ A_i= {\rm const~~for~} i=1,2,3, 
\end{equation}          
\noindent where $F_1,F_2,F_3$ represent the radial, transverse and normal
component of the nongravitational force, respectively.  The analytical function
$g(r)$ simulates the ice sublimation rate as a function of the heliocentric
distance $r$.

In a previous paper (\cite{Kro1} 1996, hereafter Paper I) we concluded that
the forced precession model of the rotating nonspherical cometary nucleus
adequately explains the variations of the nongravitational effects observed in
the periodic comet Wirtanen and is suitable for making predictions of the future
returns. In the present paper, we continue the study of nongravitational motion
of the 46P/Wirtanen including the astrometric observations from the last
apparition.

\begin{table*}
\caption[ ]{Orbital elements and physical parameters of the nucleus for forced
precession models linking all apparitions of 46P/Wirtanen.  Angular elements
$\omega$, $\Omega$, $i$ refer to Equinox J2000.0. Parameters $A$,
$A^{I}$ and $A^{II}$ are in units of $10^{-8}$~AU/day$^2$, the precession
factor $f_p$ is in units of $10^6$~day/AU, and time shifts $\tau $, $\tau _1 $
and $\tau _2$ are in days.}
{\footnotesize{
\vspace*{0.10cm}
\begin{tabular}{|l|c|c|c|}      \hline 
 &&& \\
\multicolumn{1}{|c|}{ } & \multicolumn{1}{c|}{\bf Model 1} &
\multicolumn{1}{c|}{\bf Model 2a}  & \multicolumn{1}{c|}{\bf Model 2b}  \\
 &&& \\  \hline
 &&& \\
 & \multicolumn{3}{|c|}{{\it O r b i t a l ~~~~e l e m e n t s} }  \\
 &&&   \\
Equinox:  & \multicolumn{3}{|c|}{ {\it J}2000.0 }         \\
Epoch:    & \multicolumn{3}{|c|}{1998 01 27}            \\
          & \multicolumn{3}{|c|}{ET $=$ JD\, 2450840.5} \\
 T        & 1997 03 14.15654    & 1997 03 14.15568  & 1997 03 14.15611  \\
 $q$      & 1.06342371          & 1.06342736        & 1.06343483        \\
 $e$      & 0.65681460          & 0.65680686        & 0.65680623        \\
 $\omega$ &356\fdg 32749        &356\fdg 33091      &356\fdg 33058      \\
 $\Omega$ &~82\fdg 20129        &~82\fdg 19688      &~82\fdg 19803      \\
 $i$      &~11\fdg 72345        &~11\fdg 72330      &~11\fdg 72321      \\
 &&& \\
 & \multicolumn{3}{|c|}{{\it N o n g r a v i t a t i o n a l ~~a n d} }  \\
 & \multicolumn{3}{|c|}{{\it n u c l e a r ~~~~p a r a m e t e r s} }  \\
 &&& \\
$A$      & $+0.62615 \pm 0.00523$    &   $+0.76153\pm 0.00986$  &      -----     \\
$A^{I}$  &        -----              &          -----           & $+0.80188\pm 0.00873$  \\
$A^{II}$ &        -----              &          -----           & $+0.67797\pm 0.00722$  \\
$\eta$   & ~21\fdg 02 $\pm$ 0\fdg 42 & 15\fdg 90 $\pm$ 0\fdg 58 & 15\fdg 48 $\pm$0\fdg 66  \\
$I_0$    & 136\fdg 00 $\pm$ 1\fdg 13 &154\fdg 64 $\pm$ 0\fdg 52 &140\fdg 83 $\pm$ 1\fdg 28 \\
$\phi_0$ & 259\fdg 60 $\pm$ 3\fdg 82 & 23\fdg 00 $\pm$ 8\fdg 98 &318\fdg 32 $\pm$ 1\fdg 82 \\
$f_p$    & $ 0.62528 \pm 0.02536$    & $+1.7461  \pm 0.1242 $   & $+1.2337  \pm 0.0694 $  \\
$s$      & $ 0.33419 \pm 0.01318$    & $+0.12148 \pm 0.04720 $  & $+0.10190 \pm 0.03421$  \\
 &&&   \\
$\tau   $ & --------               &       -----            &  $-23.476 \pm 1.247 $  \\
$\tau _1$ & --------               & $-31.623 \pm 1.691 $   &       -----            \\
$\tau _2$ & --------               & $-11.236 \pm 0.529 $   &       -----            \\
 &&&   \\
Res       & 1\farcs 82             & 1\farcs 68             &  1\farcs 59 \\
 &&&   \\  \hline
\end{tabular}
}}
\end{table*}

A forced precession of the spin axis is caused by the torque which arises when
the vector of the jet force (exerted by the outgassing on the nucleus) does not
pass through the mass center of the nonspherical nucleus. The precession rate
is a function of the nucleus orientation, the lag angle $\eta$ of the maximum
outgassing behind the subsolar meridian, the modulus of the reactive force, the
nucleus oblateness $s$ and the precession factor $f_p$ which depends on 
rotation period and  nucleus size (\cite{Sek1} 1984, 1988). The orientation of
the nucleus is defined by two angles: the obliquity $I$ of the orbital plane to
the cometary equator and the longitude $\phi$ of the Sun at perihelion. The
nongravitational force components (which vary along the orbit) are related to the
angular parameters
$\eta$, $I$, $\phi$ of the rotating nucleus by:
$$       F_i(t) = A\cdot C_i\left( \eta, I(t), v(t)+\phi (t)\right)\cdot g(r),
~~~{\rm for}~ i=1,2,3,$$
\noindent where $A=(F_1^2+F_2^2+F_3^2)^{1/2}/g(r)$, $v$ is the true anomaly of
the comet and $C_i$ denote the direction cosines which are time-dependent due
the orbital and precessional motion of the comet. We adopt Marsden's water
production curve, $g(r)$, because of its analytical form which is useful for
orbital computations. Expressions for time variations of the angles $I$ and
$\phi$, the precession rate, and the direction cosines of the nongravitational
force are given by \cite{Kro2} (1998).

The values of five precessional parameters: $A$, $\eta$, $I_0$, $\phi _0$, $f_p$
were derived in Paper I; the sixth  -- oblateness of the comet nucleus --
could not be determined because of poor observational material which then
consisted of 67 positional observations covering the period 1948--1991.  Since
that analysis, the new, high-quality data have become available from the latest
apparition 1995--1997. This allows us to refine the previous forced precession
model. The aim of this paper is to obtain the best model of the comet by linking
all the astrometric observations made during the eight recorded appearances of
the comet.

\section{Observational Material}

\noindent The distribution of the 67 observations of Comet 46P/Wirtanen carried
out before 1995 is given in Table 1 of the Paper I. At present, 247 positional
observations (494 residuals) are available from the recent apparition covering
the time interval of 26~June~1995 -- 30~December~1997.  Since the observations
are distributed highly nonuniformly over the apparitions, we replaced more than
two observations of the same day by one average comet position (so-called
normal place), which yields 300 residuals for the last return of the comet.
Combining mean residuals 1\farcs 96, 1\farcs 67, 1\farcs 52, 2\farcs 29,
2\farcs 44, 1\farcs 97 and 1\farcs 01 calculated separately for apparitions in
1948, 1954, 1961, 1967--1975 (two apparitions), 1986, 1991 and 1996,
respectively, we obtained an {\it a priori} mean residual  (\cite{Biel} 1991)
of 1\farcs 38.

\section{Forced precession model of the rotating nonspherical cometary nucleus}

\noindent
The equations of the comet motion have been integrated numerically by the
recurrent power series method (\cite{Sit1} 1979, 1984), taking into account all
the planetary perturbations. A detailed description of the forced precession model
is given in Paper I. The first set of equations contained six unknown
parameters $A$, $\eta$, $I_0$, $\phi_0$, $f_p$ and $s$  which were determined
simultaneously with six orbital elements from the observational equations by an
iterative least squares process. This solution is given in the first column of
Table~1 (hereafter Model~I). The positive value of $s$ implies that --
according to our model -- 46P/Wirtanen rotates on its shorter
axis.

\noindent However, a comparison of the mean residual of this solution with the {\it a priori} mean
residual shows that a better solution is required for the motion of the comet. Therefore, we no longer make the assumption that the activity reached
a maximum exactly at perihelion. We have found (from our orbital calculations)
that between the apparitions of 1986 and of 1991 (one revolution after the second
closest approach of the comet to Jupiter) something changed in the cometary
activity and/or in the geometry of the sublimation area on the nucleus surface.
These changes can be modelled by setting the discontinuity of parameters $A$
and/or $\tau$ close to the moment of aphelion passage between 1986 and 1991:
$$
   A = \left\{ \matrix{   A ^{I}  \mbox{~for $t<1989.0$~}\cr 
                          A ^{II} \mbox{~for $t\geq 1989.0$~}\cr} \right.
\mbox{~and/or~}
\tau = \left\{ \matrix{  \tau _1  \mbox{~for $t<1989.0$}\cr
                         \tau _2  \mbox{~for $t\geq 1989.0$}\cr}\right.
$$
\noindent where parameter $\tau$ represents a time shift of the maximum value
of $g(r)$ with respect to the perihelion time. Thus, the function $g(r)$ is
then replaced by $g(r')$, where $r'=r(t-\tau)$ (\cite{Sek2} 1988, \cite{Sit}
1994).

\begin{figure}
\vfill
\epsfig{file=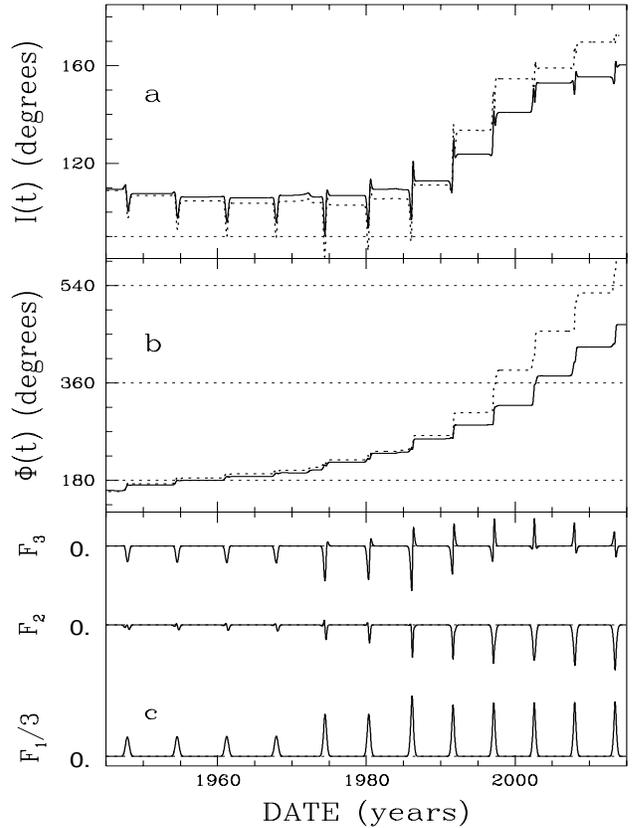,height=12.0cm,width=8.8cm}
%
\caption[ ]{Temporal variation of the angle $I$ -- {\it a},  $\phi$ -- {\it b},
and components $F_1, F_2, F_3$ of the nongravitational force {\bf F} -- {\it
c}, for Comet 46P/Wirtanen due to the spin-axis forced precession of the comet's
nucleus.  Solid curves represent Model~2b and dashed curves -- Model~2a. Dotted
horizontal lines in {\it b} indicate moments when the sunlit pole of nucleus in
perihelion changes from 'northern' to  'southern' and vice versa.}
\label{Fig1}
\end {figure}

However, it turned out that due to the relatively poor observational material
(especially that collected before 1995), only two additional parameters could be
numerically determined. Thus, we were limited to introduce the discontinuity of
$\tau$ and $A$ in two separate models, denoted as Model~2a and Model~2b respectively in
Table~1.

\noindent From this reason, the final two solutions are described by eight
parameters: \\
\noindent $A$, ~~~~~~$\eta$, $I_0$, $\phi _0$, $f_p$, $s$, $\tau _1$ 
$\tau _2$ ~~(Model 2a), \\
\noindent and \\
\noindent $A^{I}$, $A^{II}$, $\eta$, $I_0$, $\phi _0$, $f_p$, $s$, $\tau $ 
~~~~~(Model 2b), \\
which were determined from the observational equations along with the orbital
elements. These solutions, characterized by mean residuals of 1\farcs 68 and
1\farcs 59, are given in Table~1. Both models give a prediction of perihelion
passage in February 2008 (the last  before the ROSETTA rendez-vous withthe  comet)
with a time dispersion of 0.038~day.

The motion of the nucleus rotation axis represented by angles $I$ and $\phi$
and the qualitative variations of the nongravitational force components $F_1,
F_2, F_3$, acting on the comet during its successive returns to the Sun are
presented in Fig.~1 for Model~2b (solid curves). Dashed curves in the
same figure represent the variation of $I$ and $\phi$ for Model~2a; the variations of
$F_1, F_2, F_3$ are undistinguishable from those for Model~2b. This means that the
same character of the nongravitational accelerations can be interpreted by two
different models of cometary activity. The time variations of $I$, $\phi$ and $F_1,
F_2, F_3$ are a consequence of the forced precession and orbital changes caused
by two close approaches to Jupiter in 1972 and 1984. These close encounters
with Jupiter reduced the perihelion distance from 1.63~AU in 1947 to 1.08~AU in
1986, which caused an increase of the nongravitational force during this period
(Fig.~1c).  Both solutions imply a retrograde rotation of the cometary nucleus
($I>$90\degr). This is in agreement with the Jorda \& Rickman (1995) conclusion
from the positive lightcurve asymmetry combined with the negative rate of comet
orbital period change. 

\begin{table*}
\caption[ ]{Evolution of orbits of 46P/Wirtanen according to Model~2b given in
Table~1; perihelion distance $q$ is in AU, period $P$ is in years, angular
parameters $\omega$,  $\Omega$, $i$ referto Equinox J2000.0.}
\vspace*{0.10cm}
{\footnotesize{
\begin{tabular}{cccccccccc}      \hline 
 &&&&&&&&& \\
  T     &  q & e &  P &  $\omega$ & $\Omega$ & $i$ & $I$ & $\phi$ &  Epoch \\
 &&&&&&&&& \\  \hline
 &&&&&&&&& \\
1997 03 14.14929 & 1.06376188 & 0.65673107 & 5.46 & 356\fdg 34091 &  82\fdg 20593 &  11\fdg 72249 &
140\fdg 18 & 309\fdg 42 & 1997 03 13 \\
2002 08 26.63663 & 1.05876883 & 0.65780736 & 5.44 & 356\fdg 39605 &  82\fdg 17493 &  11\fdg 73837 &
145\fdg 09 & 357\fdg 62 & 2002 09 03 \\
2008 02 02.13771 & 1.05750021 & 0.65807737 & 5.44 & 356\fdg 34611 &  82\fdg 17490 &  11\fdg 73953 & 
149\fdg 17 & ~38\fdg 25 & 2008 01 15 \\
2013 07 08.64672 & 1.05209806 & 0.65923863 & 5.42 & 356\fdg 33829 &  82\fdg 16315 &  11\fdg 75731 & 
161\fdg 23 & ~85\fdg 46 & 2013 07 07 \\ 
&&&&&&&&&  \\  \hline
\end{tabular}
}}
\end{table*}

\section{Discussion}

\noindent
The present precession models impose some constraints on the physical properties of
the nucleus of Comet 46P/Wirtanen. Using these solutions it is possible to
calculate values of Sekanina's (1984) torque factor $f_{tor}=f_p/s$. Using
this parameter we are able to calculate the value of the $P_{rot}/R_a$ ratio from
the formula: $P_{rot}/R_a=4\pi \cdot f_p /(5s)$, where the equatorial radius,
$R_a$, is related to the oblateness and the polar radius, $R_b$: $R_b/R_a=1-s$. Our
models give:

\begin{equation}
 P_{rot}/R_a   = 
\left\{ \begin{array}{rl} 
0.75\pm 0.01    & \mbox{for Model~1}   \\ 
5.80\pm 1.85    & \mbox{for Model~2a}  \\ 
4.88\pm 1.36    & \mbox{for Model~2b}  \\ 
\end{array} \right.   
\end{equation}

\noindent where $P_{rot}$ is in hours and $R_a$ in kilometers. Taking into
account observations of the nucleus radius of 46P/Wirtanen it is possible to verify
our models.

\noindent From several estimates of the nucleus radius of Comet Wirtanen it
appears that this comet possesses one of the smallest nuclei detected so far.
From photographic magnitudes \cite{Jorda} (1995) obtained a radius of the nucleus
''of less than 2 km, probably 1.5--1.8~km''. This is in agreement with
\cite{Almeida} (1997), who derived a lower limit of the nucleus radius of 1.0~km
from visual magnitudes of the comet. However, systematically smaller sizes of
the nucleus are reported from CCD photometry. CCD observations done by
\cite{Boe} (1997) give an upper limit of 0.8~km and \cite{Lamy} (1998) obtained
a mean effective radius of $0.60\pm0.02$~km. Such small radii imply that a large
fraction of the surface of the nucleus is active. The active area of 1.80 km$^2$
obtained by \cite{Farnham} (1998) covers about 7\% of the sunlit surface area
for a radius of 2~km, 45\% for a radius of 0.8~km, and rises to more than 80\%
for a radius of 0.60~km.

\noindent Assuming that the nuclear radius of 46P/Wirtanen most likely is in the
range 0.5--2.0~km we obtain from Eq.~(2):
$$\begin{array}{llll}  
 R_\mathrm{eff} \mbox{~[km]~}:\mbox{~~~~~~~~} 0.5 \mbox{~~~~~~~} & 1.0 \mbox{~~~~~~~} & 2.0 & \mbox{~~~~~~~~~~~~~~~~~~~~~~~~~~~~~} \\
\end{array} $$
$$P_{rot} \mbox{~[hrs]} = 
\left\{ \begin{array}{llll} 
.42 \pm .01  & .83 \pm .01 & 1.66\pm .02 & \mbox{for Mod 1}  \\ 
3.0 \pm 0.9  & 6.0 \pm 1.8 &12.0 \pm 3.7 & \mbox{for Mod 2a} \\ 
2.5 \pm 0.7  & 5.0 \pm 1.4 &10.0 \pm 2.8 & \mbox{for Mod 2b} \\ 
\end{array} \right. $$ 
 
\noindent where $R_\mathrm{eff}$ is the effective radius of the nucleus. Thus,
Model~1 gives very small rotational periods (fast rotation). Assuming that the mean
density of the  cometary material is equal to 1~g/cm$^3$ and the comet radius is
equal to 1~km, one finds that  the gravitational force at the nucleus surface 
is almost 20 times smaller than the centrifugal force that should cause rotational
splitting of the nucleus. Alternatively, Model~1 requires the comet radius
to be greater than 4~km.

\noindent Based on observations made at intermediate heliocentric
distances and assuming a double-peaked light curve, \cite{Meech} (1997) found a
possible rotational period of about 7.6 hours. \cite{Lamy} (1998) derived a
double peaked light curve with a rotational period of 6.0$\pm$0.3 hr. Models
2a and 2b are in agreement with these results. Assuming 6~hr for the rotational
period our Models~2a and 2b give a radius in the range of 0.7--1.5~km and of
0.9--1.7~km, respectively.

\noindent From the range of variation of the light curve \cite{Lamy} (1998)
found that the ratio of the axes must be greater than 1.20 for the assumed
ellipsoidal body. Our values of 1.14$\pm$0.05 (Model~2a) and 1.11$\pm$0.05
(Model~2b) for the $R_a/R_b$ ratio are only slightly below their lower
limit. Thus, we concluded that Models~2a and 2b are to be preferred over  Model~1
because of the smaller mean residual, and the reasonable prediction of the $P_{rot}/R_a$
ratio.

However, not all observational facts are explained by our models. The negative
values of time shifts $\tau$ (Model 2b) and $\tau_1$, $\tau_2$ (Model 2a)
suggest that the gas production rate peaks before the perihelion passage. This
is in contradiction with the positive light curve asymmetry given in
\cite{Jorda} (1995) assuming that the jet force is produced only by sublimation
directly from the nuclear surface. This question was also pointed out by
\cite{Rickman} (1998). To explain this problem, they discussed a model of
46P/Wirtanen in which the nucleus is subject to fragmentation along with
sublimation. The icy chunks which break from the surface  are expelled
by the drag of the outflowing gas from the pores and propelled by a jet
force due to sublimation. They argued that if the outgassing from icy
circumnuclear material is predominant then it gives rise to the postperihelion
brightness. Thus, it is possible that the gas production curve related to the
nongravitational effects has no correspondence to the postperihelion shift of
 lightcurve maximum observed for 46P/Wirtanen. Furthermore, it is possible
that the discontinuity in global activity obtained from orbital calculations
would  not be detected in the visible brightness of the comet. More studies are
needed to judge whether the fragmentation hypothesis is true.

For all of these reasons, we conclude that our forced precession model for the
rotating comet nucleus (Model 2b or 2a) reproduces -- with very good
precision -- the nongravitational motion of 46P/Wirtanen during an almost
fifty-year interval and gives nucleus parameters (e.g.  size and rotational
period) consistent with the available photometric observations as well. Future
evolution of the orbital elements and orientation of the comet spin-axis is
presented in Table~2 for  Model~2b since this model represents the orbital
motion with the smallest mean residual. One should note that the
nongravitational forces have a significant influence on the cometary position
even during a single apparition of the comet. Thus, these effects -- even within
a single apparition -- must be included into any accurate orbit calculations.

\begin{acknowledgements}

We are deeply indebted to Professor Grzegorz Sitarski for numerous fruitful
discussions on various aspects of our investigations. We wish also to thank
Dr. Nalin Samarasinha as the Referee for his many helpful criticism and
constructive comments on the earlier version of this paper.

\noindent This work was supported by the Polish Committee of Scientific
Research (the KBN grant 2.P03D.002.09)
\end{acknowledgements}

\end{document}